\documentclass[sigconf, preprint]{acmart}
\AtBeginDocument{%
  \providecommand\BibTeX{{%
    \normalfont B\kern-0.5em{\scshape i\kern-0.25em b}\kern-0.8em\TeX}}}

\setcopyright{none}
\acmConference[preprint for VRST '22]{preprint for 28th ACM Symposium on Virtual Reality Software and Technology}{November 29-December 1, 2022}{Tsukuba, Japan}

\acmSubmissionID{1585}

\usepackage{microtype}                 
\PassOptionsToPackage{warn}{textcomp}  
\usepackage{textcomp}                  
\usepackage{amsfonts}
\usepackage{mathtools}

\usepackage{bm}
\usepackage{enumitem}

\usepackage{colortbl}

\usepackage{url}

\usepackage{xspace}
\newcommand{\eg}{e.g.\@\xspace} 

\setlist[itemize]{leftmargin=4.0mm}

\usepackage{color}
\definecolor{Orange}{rgb}{1,0.5,0}

\definecolor{DarkGreen}{rgb}{0,0.5,0}
\definecolor{Purple}{rgb}{0.7,0,0.7}
\definecolor{Blue}{rgb}{0.2,0.2,0.8}
\definecolor{Red}{rgb}{1.0,0.0,0.0}
\definecolor{Brown}{rgb}{0.7,0.4,0.1}

\begin{document}

\title{NeARportation: A Remote Real-time Neural Rendering Framework}


\author{Yuichi Hiroi}
\affiliation{%
  \institution{Sony CSL Kyoto}
  \institution{The University of Tokyo}
  \city{Kyoto / Tokyo}
  \country{Japan}  
}
\email{yhiroi@g.ecc.u-tokyo.ac.jp}

\author{Yuta Itoh}
\affiliation{%
  \institution{The University of Tokyo}
  \institution{RIKEN AIP} 
  \city{Tokyo}
  \country{Japan}
}
\email{yuta.itoh@iii.u-tokyo.ac.jp}

\author{Jun Rekimoto}
\affiliation{%
  \institution{Sony CSL Kyoto}
  \institution{The University of Tokyo}
  \city{Kyoto / Tokyo}
  \country{Japan}
}
\email{rekimoto@acm.org}


\begin{abstract}
While presenting a photorealistic appearance plays a major role in immersion in Augmented Virtuality environment, displaying that of real objects remains a challenge. 
Recent developments in photogrammetry have facilitated the incorporation of real objects into virtual space. 
However, reproducing complex appearances, such as subsurface scattering and transparency, still requires a dedicated environment for measurement and possesses a trade-off between rendering quality and frame rate. 

Our NeARportation framework combines server–client bidirectional communication and neural rendering to resolve these trade-offs. 
Neural rendering on the server receives the client's head posture and generates a novel-view image with realistic appearance reproduction that is streamed onto the client's display.
By applying our framework to a stereoscopic display, we confirm that it can display a high-fidelity appearance on full-HD stereo videos at 35-40 frames per second (fps) according to the user's head motion.
\end{abstract}

\ccsdesc[500]{Human-centered computing~Mixed / augmented reality}

\keywords{augmented virtuality, appearance reproduction, real-time rendering, neural rendering, remote rendering}


\maketitle

\vspace{-2mm}
\section{Introduction}
Photorealistic reproduction of the appearance and shape of real-world objects in virtual form is important in enhancing immersion and value judgments in a wide range of VR and AR applications. 
However, it remains difficult to reproduce the photorealistic appearance aspects of real objects, such as gloss, transparency, and subsurface scattering, on displays.

Recent mobile photogrammetry technology allows nonexperts to measure the shape and color of real objects and easily import them into virtual space.
However, these photogrammetry technologies assume that the surface is constant in radiance from any view angle. Thus, the surfaces of the reproduced subjects often possess a plaster-like texture. 
Researchers often utilize special hardware such as dome-shaped setups with structural light and multiple cameras~\cite{debevec2000lightstage, ma2007lightstage} to measure the complex, viewpoint-dependent light scattering on its surface, which is both costly and inaccessible.
Moreover, even if complex light-scattering models can be measured, high-quality rendering models that can represent light scattering, such as ray tracing, create a trade-off between image quality and frame rate.

To resolve these trade-offs in measurement and rendering, we propose NeARpotation, a framework that combines server-client bidirectional communication and neural rendering. 
NeARportation interactively displays the photorealistic appearances of real objects measured from readily available color images.
In this framework, an AR/VR device communicates with a remote GPU server running neural radiance fields (NeRF)~\cite{mildenhall2020nerf} to generate and transmit images according to the user's viewpoint. 
We apply our framework to a naked-eye stereoscopic display and show that our framework can render photorealistic full-HD images at around 35 fps from stereoscopic viewpoints, tracking the user's viewpoint position.

To the best of our knowledge, this is the first demonstration that neural rendering has the potential to interactively benefit AR/VR applications at reasonable frame rates and latencies on high-end display devices.
Our major contributions include the following:
\begin{itemize}
    \item Proposing NeARportation, a remote, real-time neural rendering framework for displaying photorealistic appearance in AR/VR devices.
    \item Demonstrating a proof-of-concept system of the NeARportation framework on a naked-eye stereoscopic display.
    \item Evaluating the reproducibility of appearance, achievable frame rate, and motion-to-photon latency in stereoscopic rendering.
    \item Providing discussions and future research directions for the current NeARportation framework.
\end{itemize}
\vspace{-2mm}

\section{Related Work}
Reproducing the photorealistic appearance of real objects has long been a research challenge. 
Conventional rendering approaches explicitly estimate a reflective model of an object's surface in reproducing a photorealistic appearance.
Hardware approaches include domed structural lighting and cameras~\cite{debevec2000lightstage, ma2007lightstage, broxton2020immersive} or light field cameras~\cite{tim2017hallelujah, wang2017lightfield, overbeck2018lightfield}, yet these dedicated camera systems maintain extensive measurement and reproduction costs. 
While deep neural network-based methods can estimate reflection models from color cameras~\cite{asselin2020svbrdf, vecchio2021surfacenet, alen2022svbrdf}, it remains difficult to generalize explicit reflection estimations to various object appearances.

Real-time rendering is a continuous research theme, mainly in game graphics~\cite{haines2018real}. 
When rendering a realistic appearance, ray casting in the rendering process can dramatically improve the expressive power of the image.
However, state-of-the-art full ray-casting real-time rendering requires over 11 GB of GPU memory~\cite{huang2020omniverse}, making it unsuitable for AR/VR devices with relatively low computing power and less energy for extended use. 

NeRF~\cite{mildenhall2020nerf} employs volume rendering to compose new view images and can handle light attenuation and transparent objects.
NeRF has developed remarkably within the past few years~\cite{xie2022neuralfields}, and new models have further extended appearance representation~\cite{Wizadwongsa2021NeX, wu2021diver, suhail2021lightfield} and realized real-time inference~\cite{garbin2021fastnerf, wang2022fourier, hu2022efficientnerf, sun22voxelgrid, mueller2022instant}.
Our framework aims to achieve multiview, realistic appearance, and high-frame rate rendering in AR/VR by leveraging both the real-time performance and appearance representation of NeRF.

While recent NeRF architectures can generate various appearances in real time for AR/VR applications, the trade-offs of resolution, number of viewpoints, and computing resources remain unresolved.
Our framework introduces a remote rendering system~\cite{shi2015remoterendering} that facilitates the process of rendering 3D graphics when the computing power of the local device is insufficient.
In remote rendering, point cloud streaming~\cite{schwarz2019vpcc, vanderHooft2019pointcloud} and video streaming rendered by ray tracing~\cite{Gul2020lowlatency, seo2022raytrace} exist as 3D content delivery. However, these works do not fully consider the application of neural rendering in remote rendering, except for a conccurent work~\cite{kondo2022immersive} that uses remote rendering using NeRF with game engines. 
While the focus of this concurrent work is NeRF interaction in VR, we focus on stereoscopic and high-resolution presentation in the present study.

\vspace{-1mm}
\section{System Overview}\label{sec:sys-overview}
\begin{figure}[tb]
 \centering
 \includegraphics[width=\linewidth]{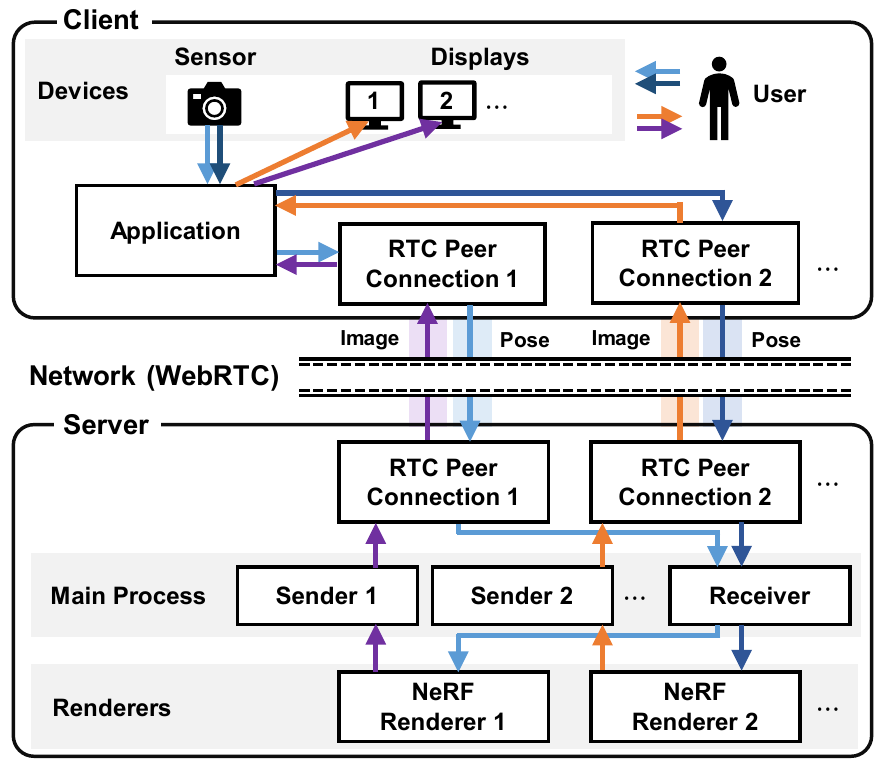}
\vspace{-3mm}
\caption{An overview of the NeARportation framework.}
\Description{The client is depicted at the top of the diagram, and the server at the bottom, both connected through the network depicted at the medium. The client consists of a sensor that detects the user's head, multiple displays, an application that integrates them, and a peer connection that transmits the information to the server. The server consists of peer connections, sender and receiver processes that control the data (head poses and rendered images), and NeRF renderers.}
\vspace{-4mm}
\label{fig:system-overview}
\end{figure}

Figure~\ref{fig:system-overview} shows a system overview diagram of NeARportation.
Our framework utilizes WebRTC-based remote rendering between a client with low computational resources (Fig~\ref{fig:system-overview}, top) and a GPU server (Fig~\ref{fig:system-overview}, bottom). 

The client consists of a sensor (\eg, a camera) to acquire the user's head pose, displays for rendered images streamed from the server, RTC peers, and an application such as a game engine.
The server consists of the main process that mediates data transmission and reception, NeRF-based renderers, and RTC peers.
The main process on the server includes a single receiver and senders equal to the number of peers.
These peers and NeRF renderers are launched with the same number of displays on the client.
Each RTC peer connects to internal processes with interprocess communication (IPC) to mediate between the processes and the P2P connection. 

Before remote rendering, the client uploads a video of the object to be reconstructed in advance to train the NeRF model, which corresponds to the design of a virtual 3D scene in conventional AR/VR applications.
Using the trained NeRF model, the server and client peers establish P2P connections.
After connecting, the server sends the client the intrinsic parameters of the camera used to capture the video.
The client sets the virtual cameras using the received parameters to reproduce the actual size of the real objects.
If there are multiple viewpoints on the client (\eg, stereopsis), the virtual cameras are set for all viewpoints.

When the client completes the configuration of the virtual cameras at every frame, the application sends the pose matrix of each viewpoint camera to the server.
Each packet received also contains labels for each viewpoint (\eg, for the left and right eye in the case of a stereoscopic display). 
The receiver on the server reads this label and distributes the pose matrix to the corresponding renderer process.

Each renderer on the server includes a buffer to store the pose matrix and a NeRF process.
The pose matrix sent from the receiver to the renderer is stored in the pose matrix buffer.
The matrix on the pose buffer is always overwritten with the latest one, and then the NeRF acquires the latest pose matrix and starts a new rendering immediately after the rendering is finished.
After the NeRF completes the image rendering, it sends the image to the sender corresponding to each virtual camera label. The sender then delivers the rendered images to the client through MediaStream on WebRTC.
Finally, the client draws the received images on the texture of a planar object that aligns exactly with each virtual camera's field of view to present them on the displays.

\section{Implementation}\label{sec:implementation}
We implement a proof-of-concept system for displaying photorealistic appearance using NeARportation by applying our framework to a stereoscopic display.

\subsection{Server Architecture}
We use a desktop machine with Ubuntu 20.04 OS, 2x Intel (R) Core (TM) i9-10920X CPU  @ 3.50 GHz, and 2x NVIDIA GeForce RTX 3090 GPUs as a server-side machine. This machine is connected to a 1-GbE wired network.

We adopt instant neural graphics primitive (instant-ngp)~\cite{mueller2022instant} for the NeRF renderer.
We launch instant-ngp instances on two different GPUs.
Each NeRF renderer process accesses instant-ngp through Python binding.

We use Momo and Ayame Labo (Shiguredo Corp.) as WebRTC clients and the signaling server, respectively. 
Momo clients encode and stream the videos from the virtual camera devices activated on the server.
We choose VP 8 as an encoding format. 
As an IPC pipeline, we use \textit{nanomsg}\footnote{https://nanomsg.org/} as a messaging library, which establishes an IPC pipeline via port-to-port TCP connections. 

\subsection{Client Architecture}
\begin{figure}[tb]
 \centering
 \includegraphics[width=\linewidth]{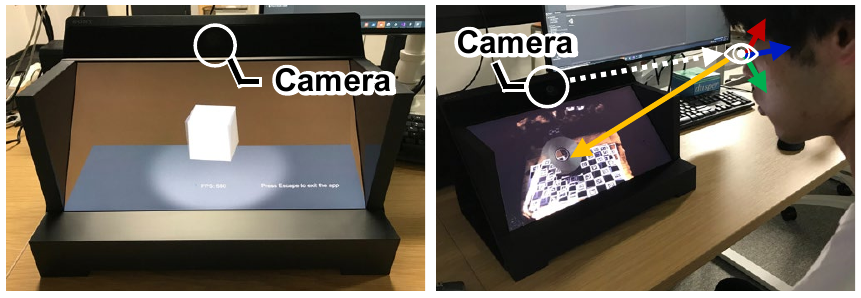}
\vspace{-3mm}
 \caption{(Left) The naked-eye stereoscopic display used in our prototype (SONY ELF-SR1). The camera above the display detects the user's head pose. (Right) A user viewing an image on a stereoscopic display. The head posture detected by the camera is synchronized with the virtual camera to show the same perspective image on display.}
 \Description{(Left) Photograph of the naked-eye stereoscopic display captured from the front. (Right) Photograph of a user looking at a naked-eye stereoscopic display placed on a table.}
\vspace{-4mm}
 \label{fig:sony-elf}
\end{figure}

We use a desktop machine with Windows 10, an AMD Ryzen 5950X CPU @ 3.40GHz, and an NVIDIA GeForce RTX 3090 GPU as a client-side machine. 
This machine is connected to a 2.4-GHz Wi-Fi network in a 1-GbE LAN environment.
The server and client are connected to different network segments.

As a naked-eye stereoscopic display, we use a Sony ELF-SR1 (Fig.~\ref{fig:sony-elf}).
The display comprises a $3840\times2160$-pixel LCD panel angled at 45$^\circ$ and a high-speed camera sensor for tracking the user's head pose.
Micro-optical lenses are installed on the front of the panel to guide two images to the left and right eyes to achieve stereoscopic viewing.
Corresponding to the training images, we use the half-resolution of this panel (full HD).

We implement the virtual application in Unity 2020.3.36f and Sony Spatial Display SDK for Unity.
The head posture detected by the sensor is immediately reflected in the anchors of the virtual cameras in the left and right eyes.
The images rendered from the virtual cameras are fused according to the microlens geometry.
We use Microsoft MixedReality-WebRTC\footnote{https://github.com/microsoft/MixedReality-WebRTC} as a WebRTC client.
The client applies the video received from the server to a planar object.

\subsection{Training Image Acquisition}
For stereopsis, a stereo image pair must use the correct disparity according to the user's interpupil distance.
Although NeRF generally uses COLMAP~\cite{schoenberger2016sfm} to roughly determine initial camera poses, COLMAP does not match the actual size of objects or their vertical orientation with reality. 
Instead, we place the ChArUCo~\cite{garrido2014automatic} marker under the object and capture training images around the object. 

We train instant-ngp in $3\times10^4$ steps with a batch size of $2^{18}$, which takes approximately 4 minutes.
When rendering, we crop only the area above the ChArUco board to improve the frame rate.

\section{Experiments}
We evaluated the reproducibility of complex appearances when displaying the captured objects and the performance of our prototype.

\subsection{Appearance Reproduction}\label{sec:eval-appear-reproduction}
\begin{figure}[tb]
 \centering
 \includegraphics[width=\linewidth]{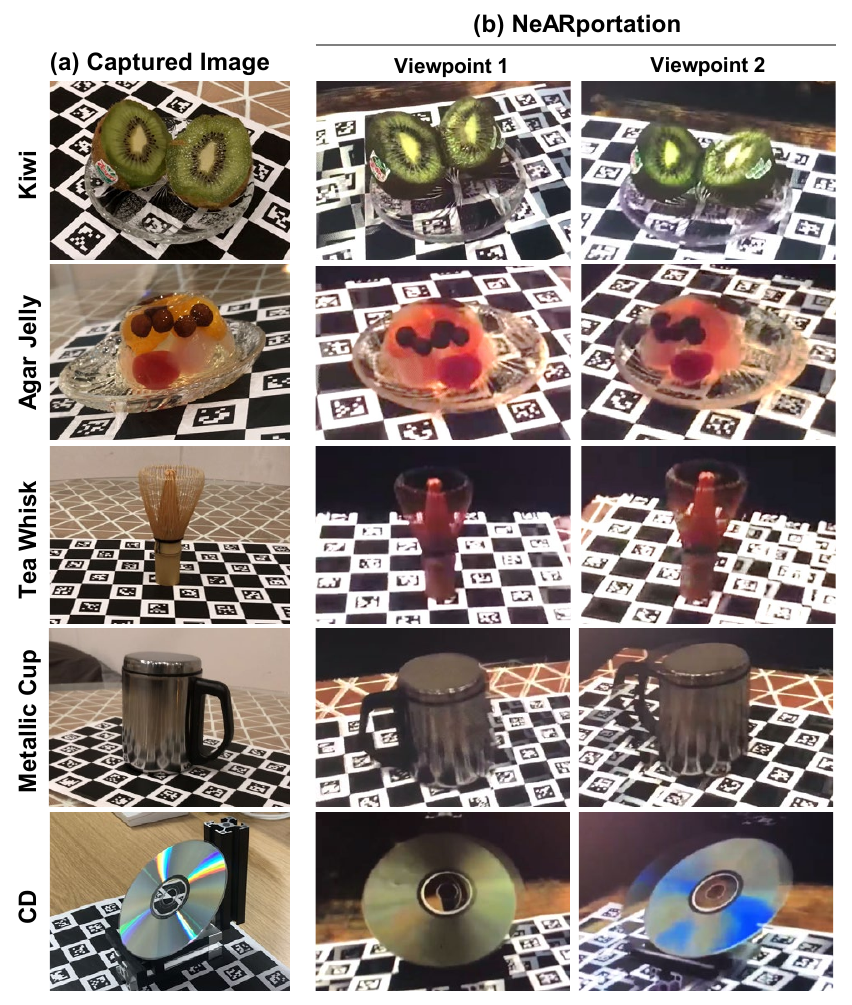}
 \caption{Qualitative performance overview of the NeARportation framework. (a) Captured images of real objects. (b) Real capture images of the reproduced appearances displayed on ELF-SR1, taken from two different viewpoints. Note that parallax artifacts appear in the images capturing the stereoscopic display because each image is taken directly with a monocular camera. This framework is interactive and is best viewed on video; thus, we urge readers to view our supplemental video.}
 \label{fig:result-displayed-images}
 \Description{The images captured from the naked-eye stereoscopic display are arranged in five rows and three columns. Images of kiwi, agar jelly, tea whisk, metallic cup, and CD are on each row. Each column contains (a) the images of the captured real objects and (b) the images of the object reproduced on the naked-eye stereoscopic display.}
 \vspace{-3mm}
\end{figure}
We captured real objects with complex appearances and displayed the reproduced stereo images with our NeARportation prototype.
To obtain training images, we first captured the object with the iPhone SE 2 for 3-5 minutes and cropped the video as images at 2 fps. 
These images were trained in instant-ngp and displayed on the application using our NeARportation framework.
Figure~\ref{fig:result-displayed-images} (a) shows examples of our training images. The objects used in the experiment and the number of training images are as follows: a kiwi cut in half (267 images), agar jelly with fruits (288 images), a tea whisk (274 images), a metallic cup (317 images), and a CD (440 images). 

Figure~\ref{fig:result-displayed-images} (b) shows the reproduced appearances.
From the 1st and 2nd rows of Fig.~\ref{fig:result-displayed-images}, we confirmed that NeARportation could interactively display the sparkling at the cross-section of a kiwi or the complex refraction in agar from novel viewpoints.
Additionally, we confirmed that our framework has sufficient reproducibility for subtle, gaze-dependent appearances following the user's head motion, such as very thin objects (3rd row), mirror reflections (4th row), and structural colors (5th row). 

\begin{figure}[tb]
 \centering
 \includegraphics[width=\linewidth]{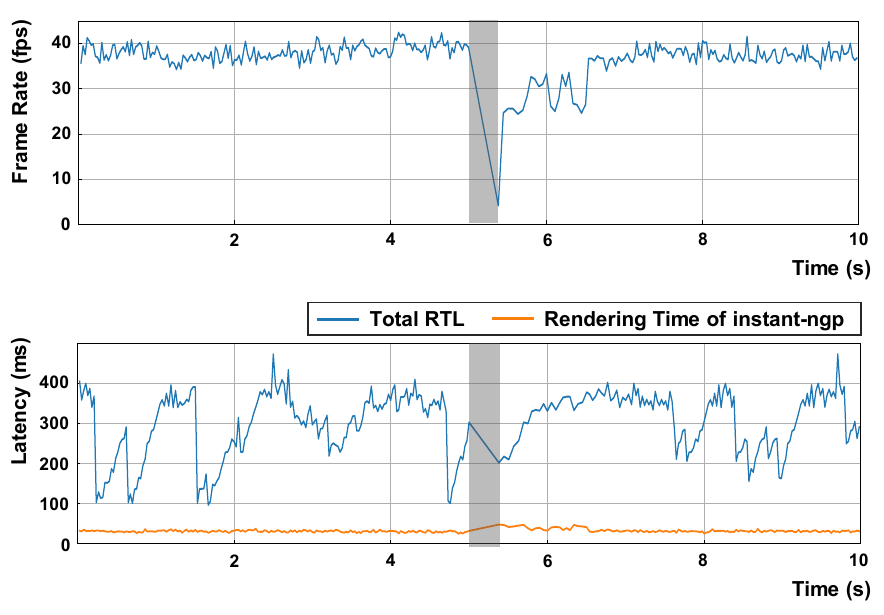}
 \caption{Transitions of (top) frame rate and (bottom) RTL and rendering latency when we keep NeARportation running for 10 seconds. The gray area represents when the application checks the communication between the client and the server at regular intervals.}
 \label{fig:result-performance}
 \Description{(top) Line graph showing frame rate (fps) from 0 to 40 on the y-axis against two increments from 0 to 10 on the x-axis (seconds); a gray area is drawn from 5.0 to 5.4 s. From 0 to 5.0 s, the line moves from 35 to 45 fps. After the communication is confirmed, the speed drops to 5 fps but recovers to 35 fps about 1.5 seconds later. (bottom) A line graph showing latency from 0 to 400 ms on the y-axis against time in 2-second increments from 0 to 10 seconds on the x-axis. Two lines are drawn: total RTL (blue) and instant-ngp rendering time (orange). The blue line moves from 100 ms to 400 ms, while the orange line stays around 30 ms.}
 \vspace{-5mm}
\end{figure}
\vspace{-1mm}
\subsection{Performance Evaluation}\label{sec:eval-performance}
Our NeARportation framework is intended to perform neural rendering interactively. To evaluate the performance toward this goal, we measure the current prototype's frame rate and round-trip latency (RTL) for 10 seconds.
To measure these metrics, we added millisecond timestamps to packets sent by the client. The server then returned the timestamp received from the client intact; at the same time, it returned the corresponding image. 
The client measured the RTL by the difference between the current time and the received timestamp and the frame rate by the timestamp interval.
We also added the rendering time of instant-ngp for each image to the returned packet to determine the percentage of NeRF rendering in the RTL.
During the measurement, the users move their heads to see the effect on rendering time.

Figure~\ref{fig:result-performance} shows the results of the performance measurement. 
In the current implementation, the application regularly checks whether P2P communication between the client and server is sustained.
Therefore, video streaming is momentarily interrupted during this period (approximately 200 ms).
Both frame rate and latency are restored to their original values after the application confirms communication with the server.

From Figure~\ref{fig:result-performance} (a), the frame rate of our prototype is around $35 \sim 40$ fps, which confirms that our prototype can provide sufficient frame rates for interactive applications such as games and videos. 
In contrast, we need further improvements to achieve the frame rates required for modern VR headsets (>90 fps)~\cite{cuervo2018creating}.
We will discuss the ways to extend our framework to higher frame rates in terms of cloud computing and improved neural network models.

From Figure~\ref{fig:result-performance} (b), the RTL of this prototype oscillates at about $100 \sim 400$ ms.
Additionally, we see that the rendering time of instant-ngp is almost constant regardless of the head position ($30 \sim 40$ ms). 
From the results, we confirmed that most of the RTL on this system depends on round-trip latency on the network.
This RTL could be improved with future network bandwidth improvements. Later, we discuss possible research directions to compensate for network latency by predicting head motion.

\section{Discussion and Future Work}
We investigate the remaining issues of the current NeARportation framework and discuss the prospects of future research directions.

\vspace{-1mm}
\paragraph{Head Motion Prediction}\label{subsec:discuss-head-predict}
An RTL exceeding 20 ms is perceived to have detrimental effects~\cite{Juraj2012virtualtexture}. However, due to the nature of our framework, which uses remote rendering, it is difficult to keep latency below 20 ms.
Compensating for this latency is currently out of scope, as this paper focuses on proving the concept for our framework.
One future research direction involves compensating for this RTL using motion prediction, such as the Kalman filter-based approaches~\cite{swafford2014dualsensor, gul2020kalman} or a time-sequential neural network model~\cite{wu2019futurepose}. 

\vspace{-1mm}
\paragraph{Improving Frame Rate}\label{subsec:discuss-distrib-render}
Reducing rendering time is essential for improving the frame rate.
A simple approach to reduce rendering time is distributed rendering with multiple GPUs, in which poses stored in a buffer are split and rendered on multiple GPUs.
Another future research direction involves a novel NeRF network, which can render images from multiple poses simultaneously on a single GPU. In this direction, an efficient method of caching scenes between viewpoints~\cite{hu2022efficientnerf} can be applied.

\vspace{-1mm}
\paragraph{Perceptive Alignment with Real Environment}
The appearance reproduced by our framework will be improved by blending the image with the environment in which the display is placed, such as color and dynamic range correction, depth-of-field reproduction, and relighting.
Although various works regarding relighting in NeRF exist~\cite{Srinivasan2021NeRV, boss2021nerd, zhang2022invrender}, to the best of our knowledge, no model has been proposed to perform it in real time.

\section{Conclusion}
We proposed NeARportation, a framework for displaying photorealistic appearance with remote rendering and a neural renderer.
The proof-of-concept system on a face-tracking stereoscopic display showed that our framework can interactively present objects' complex appearances (e.g., specular reflection, structural color, and subsurface scattering) from the easily available color images.
We also described the prospects for improving such remote rendering for photorealistic appearance.
Our work aims to inspire the community along this framework to further improve the system, such as higher frame rate and more accurate head motion tracking.
We believe this work opens the door to AR/VR experiences that are \textit{NeAR} reality.

\vspace{-1mm}
\begin{acks}
This project was partially supported by JST Moonshot R\&D Grant Number JPMJMS2012, the commissioned research by National Institute of Information and Communications Technology (NICT) Japan, JST PRESTO Grant Number JPMJPR17J2, JSPS KAKENHI Grant Number JP20H05958, JP20H04222, and JP22J01340, Japan.
\end{acks}

\bibliographystyle{ACM-Reference-Format}
\bibliography{main_revised_REVISED.bib}
\end{document}